\documentclass[%
 reprint,
superscriptaddress,
 amsmath,amssymb,
 aps,
 prl,
]{revtex4-2}

\usepackage{bbold}
\usepackage{graphicx}% Include figure files
\usepackage{dcolumn}% Align table columns on decimal point
\usepackage{bm}% bold math
\usepackage{xcolor}
\begin{document}

\title{Analyzing partially-polarized light with a photonic deep random neural network}

\author{Alessandro Petrini}
\affiliation{Institute for Complex Systems, National Research Council (ISC-CNR), 00185 Rome, Italy} 

\author{Claudio Conti}
\affiliation{Department of Physics, Sapienza University of Rome, 00185 Rome, Italy}

\author{Davide Pierangeli}
\email{davide.pierangeli@roma1.infn.it}
\affiliation{Institute for Complex Systems, National Research Council (ISC-CNR), 00185 Rome, Italy} 
\affiliation{Department of Physics, Sapienza University of Rome, 00185 Rome, Italy}

%\date{\today}~
\begin{abstract}
Optical neural networks are emerging as a powerful and versatile tool for processing optical signals directly in the optical domain with superior speed, integrability, and functionality.\linebreak
Their application to optical polarization enables neuromorphic polarization sensors, 
but their operation is limited to fully-polarized light.
Here, we demonstrate single-shot analysis of partially-polarized beams with a photonic random neural network (PRNN). The PRNN is composed of a series of optical layers implemented by a stack of scattering media and a few trainable digital nodes. \linebreak
The setup infers the degree-of-polarization and the Stokes parameters of the polarized component with precision comparable to off-the-shelf polarimeters. 
The use of several optical layers allows to enhance the accuracy, reduce the sensor size, and minimize digital costs, demonstrating the advantage of a deep optical encoder for processing polarization information. 
Our results point out photonic neural networks as fast, compact, broadband, low-cost polarimeters that are widely applicable from sensing to imaging.
\end{abstract}

\maketitle

Optical neural networks promise to drive the development of artificial intelligence
by performing accelerated and energy-efficient machine learning~\cite{Wetzstein2020, Zhou2021, Chen2024}. 
%thanks to the large bandwidth and massive parallelism of optical systems~\cite{McMahon2023}.
They also have a drastic impact in photonics as novel devices to detect, analyze,
and process optical information directly in the optical domain with ultralow latency
and capabilities beyond current sensors~\cite{Brunner2025}.
Recent works have showcased their functionalities and computational advantages
for optical image processing~\cite{Wetzstein2018, McMahon2023, Fang2024}, 
advanced machine vision~\cite{Fang2022, Fratalocchi2024},
biomedical tracking~\cite{Pierangeli2020},
terahertz inspection~\cite{Ozcan2023_1}, 
optical communication in fibers~\cite{Fischer2018}, %and free space~\cite{Ozcan2023_2},
phase imaging~\cite{Ozcan2024},
image recovery through complex media~\cite{Ozcan2023_3, Gu2024, Harbin2024, Shangai2025},
and photovoltaics~\cite{DiFalco2025}.
In this context, photonic computing is driving major advances in polarimetry and polarization imaging by enabling single-shot polarization measurements without polarization optics~\cite{Pierangeli2023},
novel full-Stokes cameras~\cite{Fan2023, Vuong2023, Pierangeli2024_2}, 
polarization transformers~\cite{Ozcan2023_4},
multidimensional detectors~\cite{Li2024}
and single-photon-level polarimeters~\cite{Jezek2024}.
These neuromorphic polarization devices still operate only on fully-polarized light.
Extending their functioning to partially-polarized light is crucial
as many polarization applications involve depolarization produced by natural scenes,
laser sources, and physical processes, which require measuring and controlling
the degree-of-polarization (DOP) by fast, broadband, and integrated devices.
Photonic networks capable of a low-latency measurement of the DOP would advance systems 
from remote sensing~\cite{Talmage1986} to spectroscopy~\cite{Cassidy2004} 
and biophotonics~\cite{Campos2020}, but have not yet been reported
due to the challenge of exploiting diffraction and interference effects 
for partially-polarized light.

Here, we report polarimetry of partially-polarized light by using a photonic random neural network (PRNN). The setup uses scattering media that act as random optical layers,
an intensity camera sensor, and a trainable digital backend.
The scheme is trained on laser beams of controlled DOP generated by exploiting a spatial light modulator (SLM). Testing demonstrates single-shot measurements of both the DOP and state of polarization (SOP) with accuracy comparable to a rotating-waveplate polarimeter.
The PRNN learns and infers the DOP through the statistical features of the scattered intensity. Increasing the number of optical layers further enhances the performance,
also reducing training and digital processing costs.
Our PRNN realizes a novel fast and compact full-Stokes polarimeter that
can be easily incorporated into integrated devices and optical fibers at any wavelength,
thus enabling the measurement of the DOP in sensing and imaging systems where bulky and slow polarimeters are inapplicable.

%%%%%%%%%%%%%%%%%%%%%%%%%%%%%%%%%%%%%%%%%%%%%%%%%%%%%%%%%%%%%%%%%%%%%%%%%%%%%%%%%%%%%%%%%%%%%%%
\begin{figure*}[t] %!h
\centering
\includegraphics[width=0.9\linewidth]{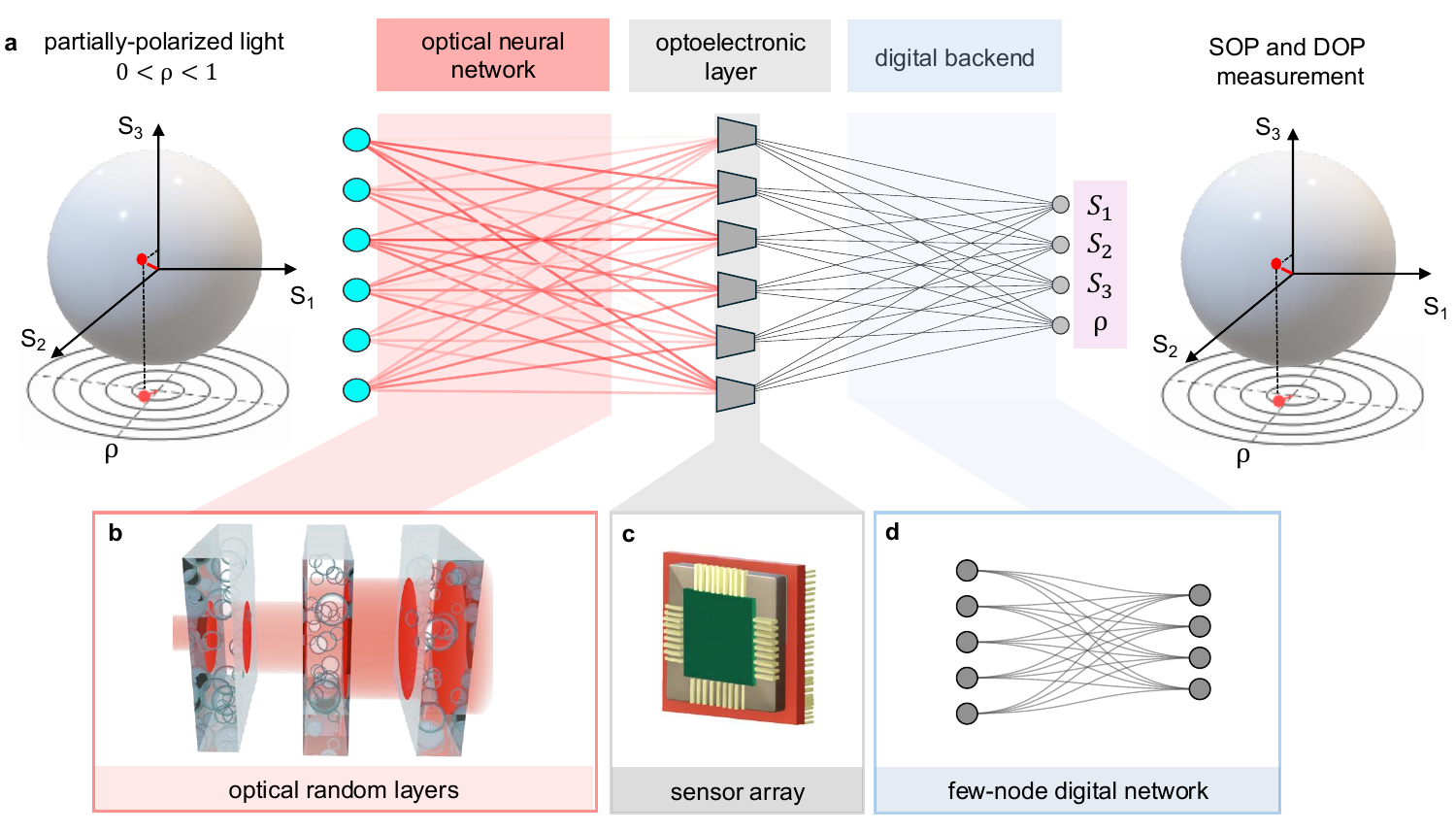} 
\caption{{\bf Polarimetry of partially-polarized light by a photonic neural network.}
{\bf a} A beam with degree-of-polarization $0<\rho<1$, 
represented as a point inside the unit Poincaré sphere, 
is first processed optically by an optical neural network. 
The transmitted intensity is detected by an array of detectors and processed by a digital backend.
The photonic network is trained to perform a single-shot measurement of the SOP and DOP
by providing the parameters $S_1, S_2, S_3$, and $\rho$.
{\bf b-d} Realization of the concept by a photonic deep random neural network.
{\bf b} The input field is self-mixed by a series of optical random layers implemented by a stack of scattering media. {\bf c} A camera sensor implements the optoelectronic layer by collecting an intensity image. {\bf d} A few-node digital neural network is trained in a supervised way to infer the SOP and DOP from the detected intensities.
}
%\vspace{-0.2cm}
\label{fig1}
\end{figure*}
%%%%%%%%%%%%%%%%%%%%%%%%%%%%%%%%%%%%%%%%%%%%%%%%%%%%%%%%%%%%%%%%%%%%%%%%%%%%%%%%%%%%%%%%%%%%%%%

\section{Photonic neural network polarimetry}
 
The concept of analyzing partially-polarized light by a photonic neural network is illustrated in Fig. \ref{fig1}a. The optical field to analyze, represented as a state lying inside the unit Poincaré sphere with DOP given by the radial distance $0<\rho<1$, is processed by an optical neural network which mixes in space or time the field components. 
This optical frontend can be constructed by diffractive neural networks,
integrated devices, metasurfaces, or any optical material that has a polarization-dependent response.
Regardless of its implementation, 
the optical network is the key element that enables polarization analysis.
In the absence of an optical frontend, 
the DOP and SOP cannot be measured by any computational methods by using the beam intensity alone. 
After the optical operation, an array of photodetectors collects the transmitted intensity, 
forming an optoelectronic layer that inputs a digital backend.
The photonic network is trained to output the Stokes parameters $S_1, S_2, S_3$ and ~$\rho$,
characterizing the SOP of the polarized component and the DOP.
\newpage

We realize the scheme by a deep PRNN where optical random layers are implemented 
via a series of scattering materials (Fig. \ref{fig1}b).
Complex media are finding broad use to realize various computing paradigms
by implementing large-scale optical random multiplications~\cite{Saade2016, Pierangeli2021, Leonetti2024, Gigan2024, Gongora2025}. 
We exploit them as hidden random layers to process polarization information.
The approach builds on a previous work that has shown the multiple light scattering 
enables learning of fully polarized vector beams~\cite{Pierangeli2023}.
The learning principle is based on the coupling between the polarization 
and spatial degrees of freedom~\cite{Pierangeli2024}.
A scattering medium projects the input field into a spatial intensity distribution 
(speckle pattern), which is detected by a camera sensor (Fig. \ref{fig1}c) and classified.
The digital backend performing the classification is a digital neural network
(Fig. \ref{fig1}d) under supervised training. 
In the training phase, it learns to associate features within the speckle pattern 
with the corresponding input field. 
We find that using a deep optical frontend allows to reduce the digital network to a few nodes.
The PRNN directly outputs the coordinates of the input field within the Poincaré sphere.
After training, it provides a single-shot measurement of the SOP and DOP 
of an unknown partially-polarized beam.

%%%%%%%%%%%%%%%%%%%%%%%%%%%%%%%%%%%%%%%%%%%%%%%%%%%%%%%%%%%%%%%%%%%%%%%%%%%%%%%%%%%%%%%%%%%%%%%
\begin{figure*}[t]
\centering
\hspace*{-0.3cm}
\includegraphics[width=1.02\linewidth]{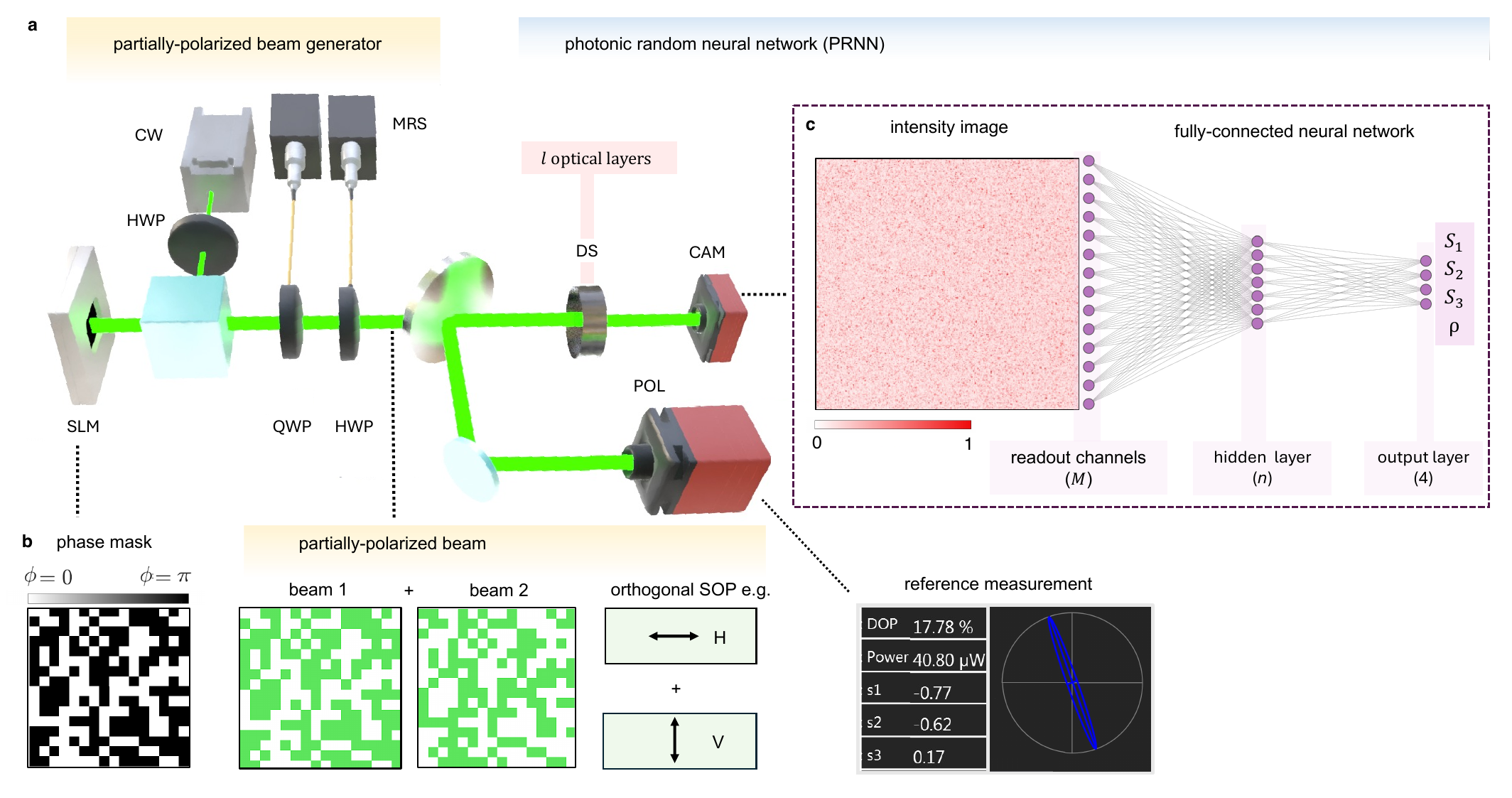}
\vspace*{-0.4cm}
\caption{{\bf Experimental setup.}
{\bf a} Scheme of the partially-polarized beam generator and the PRNN.
A stack of $l$ diffusers forms the optical deep neural network encoder. 
{\bf b}~A beam with programmable SOP and DOP is synthesized as an incoherent superposition
of two interleaved beams (beam 1, beam 2) with orthogonal SOPs (e.g. H and V) 
created by using a binary phase mask on the SLM. 
{\bf c}~Acquired intensity image and architecture of the digital backend. 
SLM spatial light modulator, DS diffuser stack, CAM intensity camera,
CW continuous-wave laser, MRS motorized rotating stages, HWP half-waveplate,
QWP quarter-waveplate, POL rotating-waveplate polarimeter.
}
\label{fig2}
\end{figure*}
%%%%%%%%%%%%%%%%%%%%%%%%%%%%%%%%%%%%%%%%%%%%%%%%%%%%%%%%%%%%%%%%%%%%%%%%%%%%%%%%%%%%%%%%%%%%%%%

\section{Learning the degree-of-polarization}

\subsection{Experimental setup}

Partially-polarized light can be generated as a spatial, temporal, or spectral, 
incoherent superposition of orthogonally polarized states~\cite{Wolf}. 
We implement and control an incoherent superposition in space by exploiting a single SLM
that operates on a fully polarized laser.
This original method represents a significant addition to the techniques for controlling simultaneously the SOP and DOP, which use temporal integration by the detector~\cite{Campos2014}, multiple beam paths~\cite{Campos2015},
or metasurfaces of fixed design~\cite{Yang2023}.
Our partially-polarized beam generator accurately produces on-demand states lying inside the Poincaré sphere at a rate of order of $10$ Hz, thus enabling rapid training of the PRNN.

The optical setup, illustrated in Fig.~\ref{fig2}a,
includes the PRNN and the generator that synthesizes beams with fully programmable SOP and DOP.
The generator is composed of a phase-only SLM (Hamamatsu X13138, $1280\times1024$ pixels, $12.5$ $\mu$m pixel pitch, $60$ Hz frame rate) sandwiched between input and output waveplates. 
The SLM screen is impinged by an expanded beam (CW laser, $\lambda = 532$ nm, $250$ mW)
with diagonal polarization set by the input half-waveplate.
The output quarter- and half-waveplate, 
oriented by motorized rotation stages at angles $\alpha$ and $\beta$, % respectively,
convert the phase delay $\phi$ imparted by the SLM 
into a SOP determined by the three parameters $(\alpha, \beta, \phi)$,
as detailed in Ref.~\cite{Pierangeli2023}.
To control also the DOP, we create two overlapping beams with orthogonal SOPs and controllable intensity by using a phase mask composed of square macropixels (blocks of pixels) 
with binary phases distributed randomly in space (see Fig.~\ref{fig2}b).
Phases $\phi$ and $\phi + \pi$ correspond to orthogonal SOPs for any $\alpha$ and $\beta$.
For example, a superposition of H and V polarizations 
is obtained for $\phi=0$, $\alpha= 45$\textdegree and $\beta=0$.
To vary the intensity of each beam of the superposition,
a fraction $p$ of the SLM macropixels is programmed with phase $\phi$, 
the remaining $1-p$ with $\phi +\pi$.
The DOP is modulated by varying $p$ and depolarized states are obtained for $p=0.5$.
We achieve full control of the SOP and DOP by programming the parameters $(\alpha, \beta, \phi, p)$.

The generated partially-polarized beams are analyzed by the PRNN,
which is made by $l$ scattering media, a CMOS intensity camera, 
and a digital neural network.
We employ a set of ground glass diffusers 
(Thorlabs N-BK7 with $120$, $200$, $600$, or $1500$ grit polishes) as scatterers.
They are added progressively in order of decreasing grit to form a stack 
to study the effect of the optical network depth on the PRNN polarimeter.
The speckle pattern is collected by the camera (Basler a2A1920-160umPRO, $1920\times1200$ pixels, $12$-bit pixel depth) positioned approximately $10$~cm apart from the stack of diffusers.
A $4\times4$ pixel binning is performed directly on the CMOS sensor,
which implements an average pooling layer performed by hardware.
The acquired images have a resolution of $M=300\times300$ pixels (Fig.~\ref{fig2}c),
with $M$ that defines the effective size of the sensor.
Reference polarization measurements of the input beam are performed by a rotating-waveplate polarimeter (PAX1000VIS, 0.25\textdegree accuracy)
placed in a secondary optical arm.  %exp time 7ms

We train the digital backend by using $N_{\rm train}$ acquired images 
and the corresponding measurements by the polarimeter as labels.
We use the fully-connected neural network shown in Fig.~\ref{fig2}c. 
It is composed of an input layer, 
where the number of input nodes coincides with the number of pixels $M$ (readout channels),
a few-node hidden layer of $n$ nodes, and four output nodes that produces 
the result of the beam analysis giving $S_1$, $S_2$, $S_3$, and $\rho$.
The hidden layer is connected to the output via batch normalization and ReLU activation. 
%This network has $8.9\times 10^6$ learnable parameters and was trained for 50 epochs using the Adam optimizer, with a piecewise learning rate schedule (initial learning rate, $0.01$, drop factor, $0.1$).
We test the PRNN on \(N_{\mathrm{test}}\) beams with unknown SOP and DOP.
Their state is inferred from the sole speckle pattern 
and it is verified by using the polarimeter as a ground-truth.

%%%%%%%%%%%%%%%%%%%%%%%%%%%%%%%%%%%%%%%%%%%%%%%%%%%%%%%%%%%%%%%%%%%%%%%%%%%%%%%%%%%%%%%%%%%%%%
\begin{figure*}[t]
\centering
\hspace*{-0.1cm}
\includegraphics[width=0.98\linewidth]{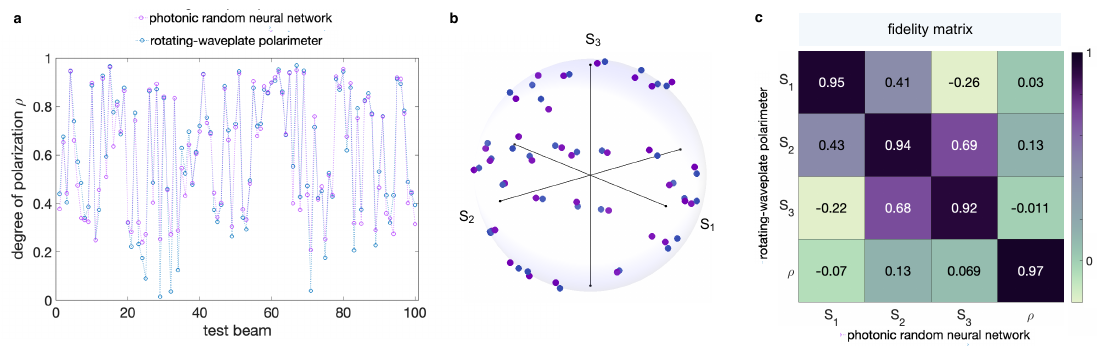}
\vspace*{-0.1cm}
\caption{{\bf Measurement of the DOP and SOP by the PRNN.}
{\bf a} DOP of $N_{test}=100$ test beams as measured by the PRNN 
(magenta dots) and the polarimeter (blue dots). 
{\bf b} Corresponding SOP of several test beams on the unit Poincaré sphere.
A mean error of ${\rm RMSE}=0.13$ quantifies the high accuracy of the inference.
{\bf c} Correlation matrix between the PRNN and polarimeter analysis.
Results are obtained with one optical layer and $n=20$ hidden nodes.}
\label{fig3}
\end{figure*}
%%%%%%%%%%%%%%%%%%%%%%%%%%%%%%%%%%%%%%%%%%%%%%%%%%%%%%%%%%%%%%%%%%%%%%%%%%%%%%%%%%%%%%%%%%%%%%%

\subsection{Testing the PRNN polarimeter}

We first demonstrate the PRNN polarimeter by using a single diffuser ($l=1$, 1500 grit).
We generate a set of partially-polarized beams with $S_1$, $S_2$, $S_3$, $\rho$
selected randomly and store the speckle images and polarimeter data,
which are split into $N_{\rm train} = 900$ training and $N_{\rm test} = 100$ testing samples.
The results on the test beams are reported in Fig. \ref{fig3}.
Figure \ref{fig3}a shows that the PRNN is able to perform single-shot measurements of the DOP.
For each beam, the SOP is also inferred accurately from the same intensity image,
even for beams with a low $\rho$.
Figure \ref{fig3}b reports on the unit Poincaré sphere the SOP of the polarized component as measured by the PRNN and polarimeter.
We evaluate the overall accuracy of the PRNN measurement by the root-mean-square-error defined
as RMSE $=\sqrt { ( \vert \rho^{\rm PRNN} - \rho^{\rm POL} \vert^2 + \sum _{i=1}^3 \vert S^{\rm PRNN}_i - S^{\rm POL}_i\vert ^2 )/4}$, where superscripts refer to values measured by the network and polarimeter, respectively.
%Note that the root-mean-square-error applied to the SOP is expected to give a value larger 
%than in other applications because the Stokes parameters are not independent, 
%i.e., a discrepancy in one of the $S_i$ reflects on the other components.
We obtain an error RMSE $=0.13\pm 0.01$. 
The good accuracy is also evidenced by the fidelity matrix (Pearson correlation) 
%person coefficient between S_i^PNNR and S_j^POL (4x4 matrix), averaged on the set
in Fig.~\ref{fig3}c. 
The fidelity of the DOP measurement reaches $97\%$,
while the $S^{\rm PRNN}_i$ are strongly correlated ($94\%$, on average) with the $S^{\rm POL}_i$. 
Cross-correlations among different Stokes components reflect their intrinsic interdependence. 
In contrast, $\rho^{\rm PRNN}$ is statistically independent of each Stokes parameter,
as evidenced by the lack of correlation.
Therefore, the PRNN operates as an effective polarimeter also in terms of correlations
within the measurement.

\subsection{Advantages of a deep optical encoder}

\begin{figure*}[t]
\centering
\hspace*{-0.2cm}
\includegraphics[width=1\linewidth]{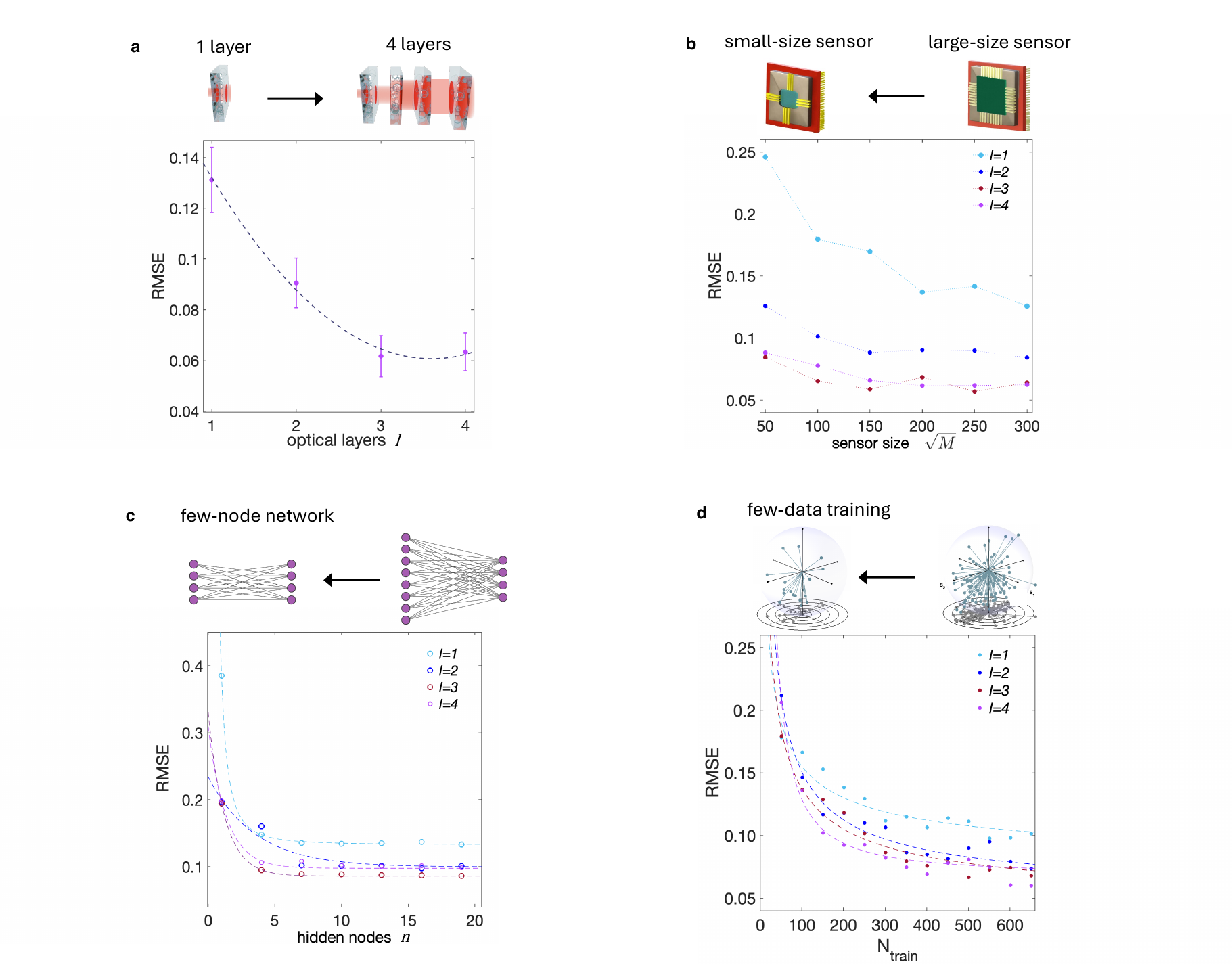}
\vspace*{-0.2cm}
\caption{{\bf Performance enhancement by a deep optical encoder.}
{\bf a} PRNN accuracy versus the number of optical random layers $l$ and 
{\bf b} varying the active area of the sensor via the number of readout channels $M$.
{\bf c} Measurement error as a function of the number of hidden nodes $n$
of the digital network and {\bf d} varying the size of the training dataset for different $l$.}
\label{fig4}
\end{figure*}

We enhance the performance of the PRNN polarimeter by augmenting the optical depth of the PRNN.
We progressively increase the number of diffusers (random optical layers)
and investigate how the depth of the optical encoder 
influences the PRNN operation in terms of accuracy, digital resources, and training costs.
Figure \ref{fig4}a reports the accuracy varying the number of layers $l$ in the stack
and keeping fixed the other hyperparameters. %$N_{train} = 900$ and  $N_{test}=100$ n=10
We observe that the SOP and DOP are inferred with enhanced accuracy as $l$ increases.
The performance saturates to a minimum error of RMSE $=0.06 \pm 0.01$ for $l \geq 3$,
which is half of the error obtained for a single diffuser.
Such a small error indicates that the PRNN reaches an accuracy comparable to the polarimeter.
%% In terms of the angles on the Poincaré sphere, the SOP is measured with an accuracy of 0.4\degree, to compere with the pm 0.5 accuracy of the polarimeter.
The enhancement is interpreted considering that cascaded diffusers 
increase the coupling between the input field and the output spatial modes.
The speckle pattern %has different statistical features and 
becomes more sensitive to changes of the input DOP and SOP, %link to memory effect
and this improve learning and inference.
In terms of random neural networks, increasing the number of optical layers corresponds to enlarging the feature space where polarization information is projected.

%%%%%%%%%%%%%%%%%%%%%%%%%%%%%%%%%%%%%%%%%%%%%%%%%%%%%%%%%%%%%%%%%%%%%%%%%%%%%%%%%%%%%%%%%%%%%%%
%%%%%%%%%%%%%%%%%%%%%%%%%%%%%%%%%%%%%%%%%%%%%%%%%%%%%%%%%%%%%%%%%%%%%%%%%%%%%%%%%%%%%%%%%%%%%%%

The enhanced accuracy allows operation with a sensor of reduced size, 
thus reducing latency and digital processing.
This additional advantage enabled by the use of a deep optical encoder
is demonstrated by decreasing the number of nodes of the optoelectronic layer.
Figure \ref{fig4}b shows the measurement error varying the readout channels $M$
for diffuser stacks of different depth $l$.
The error remains small for $l=3,4$, i.e., 
accurate measurements are achieved even in the extreme case of a sensor with very small size.
The DOP and SOP are measured with RMSE $=0.09 \pm 0.01$ 
by using a detector with only $M=50 \times 50$ pixels. 
On the contrary, the single-layer PRNN is ineffective with this small-size sensor. %n=10,tutte
Augmenting the optical depth also enables accurate measurements 
by using a reduced number of digital hidden nodes. 
The RMSE when decreasing the hidden nodes $n$ is reported in Fig.~\ref{fig4}c.
The deep PRNN ($l=3,4$) maintains a good accuracy even for a digital network of small size.
Remarkably, we achieve the SOP and DOP by using only four hidden nodes.
Moreover, empowered by several diffusers, 
the deep PRNN is also effective in condition of a few training samples.
As shown in Fig.~\ref{fig4}d, while $N_{\rm train} = 600$ still provides near-optimal performance, the PRNN with $l=4$ performs well (RMSE$=0.10 \pm 0.01$, $n=20$, $M=300\times300$) 
also when trained with only $N_{\rm train} =150$ beams.
In this case, the acquisition of the training data, i.e.,
the calibration of the PRNN polarimeter, takes less than $10$ seconds in our setup.
The results in Fig. \ref{fig4} thus demonstrate how the deep optical encoder
allows to (i) enhance the measurement accuracy (Fig.~\ref{fig4}a),
(ii) reduce the active area of the CMOS sensor (Fig.~\ref{fig4}b),
(iii) reduce the size of the digital backend (Fig.~\ref{fig4}c), and
(iv) minimize the training time (Fig.~\ref{fig4}d).
This enables partially-polarized light analysis with a minimum computational effort 
while maintaining a high precision.

\subsection{DOP learning mechanism}
 
\begin{figure}[t!]
\centering
\hspace*{-0.3cm}
\includegraphics[width=1.01\linewidth]{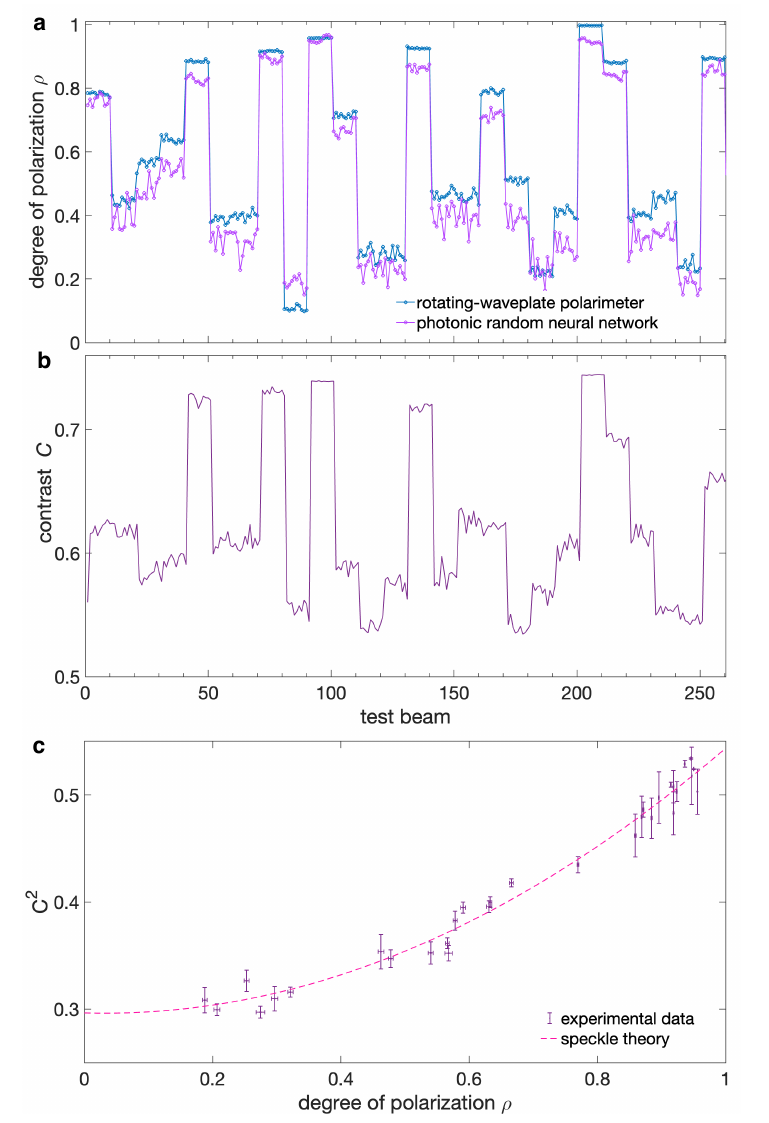}
\caption{{\bf Relation between the DOP and the speckle contrast.}
{\bf a} Measured DOP for a stream of test beams.
Each step corresponds to ten replicas with the same DOP and SOP 
but generated by different realizations  of the phase mask 
and inferred from different intensity images.
{\bf b} Corresponding image contrast.
{\bf c} $C^2$ as a function of $\rho$. Experimental data (dots) and fitting curve (line)
according to the model of partially-polarized speckles (Eq. \eqref{eq1}).
}
\label{fig5}
\end{figure}

To understand the mechanism by which the PRNN learns the DOP from the intensity data,
we investigate the properties of the speckle patterns generated by the partially-polarized beams.
For polarized light, the transmitted intensity in each spatial point depends on the incoming SOP according to the transmission tensor of the diffuser~\cite{Pierangeli2024}.
The training phase allows to interpolate this tensor, 
so the SOP is learned through its direct association with the spatial intensity features. 
However, how an unpolarized component of the beam affects the spatial intensity distribution
is less understood.
To identify a quantity within the speckle pattern that is directly related to the DOP,
we perform experiments with a designed set of beams with fixed SOP.
We generate a sequence of beams with random DOP and, for each $\rho$,
we analyze ten different realizations of the same state 
that are obtained by different spatial realizations of the phase mask on the SLM.
In this way, we study replicas with the same DOP and SOP
but associated to different speckle patterns. 
The measured DOP after training is reported in Fig.~\ref{fig5}a in comparison with the polarimeter.
%(\(N_{\rm train} = 1200, N_{\rm test} = 300, RMSE=$0.11 \pm 0.01$)
The step graph shows that the PRNN achieves the same $\rho$ for different realizations of the beam (different intensity images), i.e, the measurement is independent of the replica.
This observation reveals that the DOP is uncorrelated with the spatial positions of the speckles, which, instead, encodes information on the SOP.  
Conversely, we observe a strong correlation between the DOP and the contrast of the acquired speckle pattern $C = \sigma_I/ \langle I(x,y)\rangle$, 
where \(\langle I(x,y)\rangle\) is the mean and \(\sigma_I\) the standard deviation
of the intensity distribution in the $x$-$y$ camera plane.
As reported in Fig.~\ref{fig5}b, a lower (higher) $\rho$ corresponds to a lower (higher) speckle contrast $C$. 
The DOP thus influences the statistical properties of the transmitted intensity. 
The evidence agrees with the theory of speckle patterns produced by partially-polarized light \cite{Goodman2020}, which predicts the relation:
\begin{equation}
\label{eq1}
  C^2 = 1 + \rho^2,
\end{equation}
which applies to our case where speckle grains have a size larger than the detector area (camera pixel). We experimentally validate Eq. \eqref{eq1}.
As shown in Fig. \ref{fig5}c, 
the measured $C$ as a function of $\rho$ is well fitted by the model.
Therefore, the PRNN learns the DOP by exploiting global statistical properties of the intensity images such as their contrast.

%%%%%%%%%%%%%%%%%%%%%%%%%%%%%%%%%%%%%%%%%%%%%%%%%%%%%%%%%%%%%%%%%%%%%%%%%%%%%%%%%%%%%%%%%%%%%%%
%%%%%%%%%%%%%%%%%%%%%%%%%%%%%%%%%%%%%%%%%%%%%%%%%%%%%%%%%%%%%%%%%%%%%%%%%%%%%%%%%%%%%%%%%%%%%%%

\section{Discussion}

The PRNN polarimeter features single-shot operation and ultralow latency, 
which is set by the acquisition time of the intensity detector.
Considering the camera employed in our proof-of-concept and the exposure time of $1 \mu$s
used in experiments, images of $M=50 \times 50$ binned pixels (Fig. \ref{fig4}b)
can be acquired at nearly $700$ frames per second.
%(limited by the device link throughput limit).
%se riducessimo a 20x20 (500 exp) arriva a 1500 fps
The speed is compared with the maximum sample rate of the rotating-waveplate polarimeter 
($10$-$100$ Hz depending on the operating mode). 
Therefore, we get a intelligent polarimeter that is one order of magnitude faster
than a conventional device by using a low-cost camera 
and no polarization optics.
By employing a high-speed camera, the PRNN latency can be reduced in the $\mu$s range, 
which would enable its application to analyze laser pulses 
and to measure temporal variations of the DOP and SOP.

The device performance is comparable to the accuracy of the commercial polarimeter 
when using a deep optical encoder.
The propagation between the optical random layers improves the learning process, 
in analogy with observations for diffractive neural network classifiers \cite{Ozcan2018}.
Crucially, when increasing the number of diffusers,
the digital backend can be shrunk in sensor size and hidden nodes 
while maintaining high accuracy.
The result demonstrates the advantage of optical neural networks for sensing \cite{McMahon2023}. %Digital processing can be reduced further by introducing nonlinear optical activation within our diffuser stack.

In conclusion, we have demonstrated a photonic deep neural network 
for analyzing partially-polarized beams in a single shot. 
Our implementation combines optical processing via multiple light scattering 
with minimal digital processing by a few-node fully-connected network.
Our work introduces a novel tool for measuring the DOP and SOP
that surpasses conventional polarimetry in terms of speed and compactness.
The concept can be extended to other optical networks made by diffractive layers,
spatial light modulators, and nonlinear optical systems, 
and directly applied within a broad frequency spectrum.
Our PRNN polarimeter overcomes the latency limitation of conventional polarimeter
and can be integrated into photonic chips and optical fibers,
unleashing new possibilities in sensing, imaging, optical communication and photonic computing.

%%%%%%%%%%%%%%%%%%%%%%%%%%%%%%%%%%%%%%%%%%%%%%%%%%%%%%%%%%%%%%%%%%%%%%%%%%%%%%%%%%%%%%%%%%%%%%%

\vspace*{-0.1cm}
We acknowledge funding from the HORIZON-ERC-2024-STG LOOP No. 101162173,
HORIZON-ERC-2023-ADG HYPERSPIM No. 101139828, 
and the EU-NextGenerationEU under the National Recovery and Resilience Plan (NRRP) MUR PRIN 2022 MetaComputing No. 2022N738SA (CUP B53D23004370006), MUR PRIN 2022 PHERMIAC No. 2022597MBS 
and National Quantum Science and Technology Institute (NQSTI) MUR No. PE0000023-NQSTI.
%\newpage
%\vspace*{-0.2cm}

\end{document}